\documentclass[a4paper]{jpconf}
\usepackage{color}
\usepackage{graphicx}
\usepackage[T1]{fontenc} 
\usepackage[utf8]{inputenc}

\begin{document}
\title{Creation of a group on particle accelerator science and technology in Mexico}

\author{J. G. Contreras$^1$, M. Napsuciale$^2$}

\address{$^1$Faculty of Nuclear Sciences and Physical Engineering, 
Czech Technical University in Prague, Czech Republic}

\address{$^2$Departamento de F\'{\i}sica, Universidad de Guanajuato, Le\'on, Guanajuato, M\'exico}

\ead{jgcn@mail.cern.ch,mauro@fisica.ugto.mx}

\begin{abstract}
The story-line of the activities that have produced the first group of scientists in Mexico working on particle accelerator science and technology is briefly presented.
\end{abstract}

\section{Introduction}

In Mexico, there is a strong community of physicists working on High-Energy Physics (HEP). 
Initially, most of the practitioners were working on theoretical subjects, but by the 
early 2000s there was already  a strong group of experimentalists working on international 
collaborations at CERN, DESY and FERMILAB~\cite{Garcia:2006zza}. To many of us working in HEP, 
it seemed that the natural next step had to be to contribute also to the accelerator side 
of our field.

There were some previous attempts to develop accelerator science and technology in Mexico. 
The Institute of Physics of Guanajuato University (IFUG) founded in 1986 was conceived as 
a centre for the development of accelerator science and technology under the auspices of 
FERMILAB and other national laboratories in the USA. Unfortunately, this project did 
not succeed, in 1992 the accelerator project was closed and new groups working on 
other topics started a new era of this institute. 

An opportunity to explore again the formation of groups in this area came from the 
National Council for Science and Technology of Mexico (CONACyT), which in 2006 opened 
a call for proposals to develop large scientific and technological projects. One of the proposed 
projects was the development of accelerator science and technology with 
the ultimate end of building a light source in Mexico~\cite{Herrera2007}.

After evaluating the different proposals, CONACYT decided to create nation-wide networks 
in specific topics such that the scientist could organise themselves, work-out a set of 
priorities and develop the ideas and projects as one community. Among  the original networks 
was the Topical Network on High-Energy Physics (Red-FAE), which started working in 2008. 
One of the conclusions in the original diagnostic of the field presented during the 
First General Meeting of the Red FAE held at the city of Taxco in 2009, was the  
urgency to develop accelerator science and technology for different purposes in Mexico. 
Within this topic, one important item was the development of human resources in the different 
disciplines involved in the field. 

In this context, several groups got interested in the process, to be described below. 
On the other hand, in 2010 the then Deputy General Direction for Scientific Development of 
CONACYT, Dr. J. A. de la Pe\~na, created an ad-hoc panel with the mandate to explore the 
possibility of large scale projects in the area of accelerators. The panel recommended 
to support projects in two areas: hadron therapy and light sources~\cite{Celis2010}. 
One of the issues emphasised in the report, was the need to form human resources in all 
related fields.

The development of human resources for accelerator science and technology in Mexico 
clustered initially around two independent initiatives  to be described in the next sections.

\section{The collaboration with CERN}

One of the outcomes of the discussions leading to the proposal~\cite{Herrera2007} to build 
a light source, was that, given that most of us had little or no experience in frontier 
research in accelerators, it was clear that we had to start looking for students interested 
in this field to train them at the highest level, in order that after some time they would be the 
driving force of this type of projects. Then,  after the submission of the proposal  we started 
to look for interested students. 

In 2007, the first student willing to participate in this adventure was found. Following the 
advice of Dr. E. Elsen,  one of us (JGC) contacted Dr. Koutchouk towards the end of 2007. 
Dr. Koutchouk was finalising the preparation of an European project (EuCARD) under the FP7 
program.  After some discussion of the situation, both in Mexico and in Europe, we decided that 
the best contact to help us train this first student would be Dr. F. Zimmermann from CERN.

In 2008 the student spent the Summer at CERN working with Dr. Zimmermann and later on this 
work was the basis of his 2009 M.Sc. thesis. The student continued the collaboration with 
Dr. Zimmermann working on electron cloud effects in the LHC~
\cite{MauryCuna:2012zza,Dominguez:2013gw} and obtained his Ph. D. degree in September of 2013.
The success of this first student motivated us to send more students to CERN and Dr. Zimmermann 
helped in finding adequate supervisors for them.

From the very beginning it was clear that funding the stays of students at CERN would be 
a problem. Having no local community with experts in this field meant that there was no 
clear way to successfully submit projects to obtain grants for the students. Furthermore, 
the disparities in the cost of living in Mexico and in the Geneva area meant that the 
minimal funding required at CERN was substantially larger than the norm in Mexico. 

The funding for the initial stays was obtained from many sources. In Mexico the home institutes 
of the  students, the RedFAE and  CONACYT  contributed funds; the rest of the money was 
obtained from the  EuCARD , HELEN and E-Planet projects. 

In parallel we explored how to solve permanently the problem. Since 1998 there is a 
cooperation agreement between CERN and CONACYT. There was a possibility to write an 
addendum to this agreement to take care specifically of the collaboration in the area 
of accelerator science. This was thought to be the right way to proceed, but it took a 
long time to crystallise. Finally in 2015 such an addendum  was  signed by the representatives 
of both institutions and a formal way to request funding was finally available. The project 
was called BEAM and was de facto in operation already before the formal signature of the addendum.  

Under the auspices of BEAM three students have obtained a Ph.D. in accelerator sciences. 
In addition to the work on electron cloud effects, they have also contributed to our current knowledge 
on Crab cavities~\cite{Yee-Rendon:2014efa} and the design and construction of the 
LINAC4~\cite{Valerio-Lizarraga:2015dva}. Also three students got their M. Sc. degree from Mexican institutions, with the research work performed at CERN.

Currently there are five Mexicans working in the field of accelerator sciences at CERN. One 
is a CERN fellow, three are doing the research towards their Ph.D. and one is working 
towards his M. Sc. degree. They are working in a variety of fields covering synchrotron radiation 
at the LHC and future colliders, design testing and operation of superconducting RF cavities 
and crab cavities and optics and design of the interaction regions for the future colliders 
(TLEP, FCC-ee).

\section{The collaboration with JLAB}

During the trip back to Le\'on from the Taxco meeting in 2009, a discussion with Gerardo Moreno (first 
high energy physics experimentalist in Mexico, participant of the original IFUG project  and 
presently retired) triggered the interest in the field by one of us (M.N.). Following the 
conclusion on the need to develop accelerator science and technology, Gerardo Moreno argued that 
a starting point to this endeavour could be to involve students in projects for the development 
of compact light sources which received a boost in terms of financial support with the incorporation 
of a possibility to generate X-rays by Inverse Compton Scattering (ICS). The calculation of Compton scattering is one of the goals of every course in QED, hence highly 
standard, the novelty here is the kinematical regime when the electron is moving in whose case 
it transfers energy to the photon increasing its energy. The technical problem that has prevented 
the development of efficient machines by ICS is the luminosity of the beams and those related to the 
small size of a compact light source. 

By January of 2009, a master student at IFUG (Alejandro Castilla Loeza) got interested in the topic and 
after six months he learned QED and calculated ICS in the kinematical region of interest having 
a complete picture for the design. There were no facilities 
for the experimental part so a collaboration was established with the time-group at the National 
Centre of Metrology (CENAM) to make a feasibility demonstration experiment with old TV electron guns 
to produce X-rays from visible light.

By the end of 2009, a Mexican researcher working at the Thomas Jefferson National Laboratory 
(JLAB), Carlos Hernandez Garc\'ia, was invited by the DPyC to deliver one of the lectures at 
the XII Mexican Workshop on Particles and Fields held in Mazatl\'an. This was the opportunity 
to establish a fruitful collaboration between Mexican institutions and JLAB. At this very same 
meeting a program was designed for the Ph.D. program of Alejandro Castilla Loeza, focused on 
the development of crab cavities for the projected upgrade to 12 GeV of JLAB. After his return 
to JLAB, Carlos Hern\'andez 
got a formal invitation from JLAB to the DPyC to discuss the possibilities for collaboration with 
Mexican institutions. Since Guanajuato had already an initial collaboration with JLAB, 
one of us (M.N.) was designed by the DPyC to attend JLAB's invitation.

The visit to JLAB was a highly motivating experience and we will always be in debt 
with Carlos for the preparation in every detail of this visit which included discussions with 
the engineers in charge of every part of the accelerators at CEBAF, the living experience 
of the free electron laser group in their everyday duties, highly supportive meetings with the 
Director of the Accelerator Division ( Andrew Hutton ) and the general director of JLAB 
(Hugh E. Montgomery), as well as a general meeting to explore the modalities to support 
the development of accelerator science and technology in Mexico.
Many ideas emerged from this meeting, among them the need to organise a  periodic school 
on accelerators in Mexico, a permanent space for Mexican students in the JLAB summer 
program and in the United States Particle Accelerator School (USPAS), the 
possibility of joint Ph.D. programs between Mexican universities and Old Dominion University 
(close to JLAB) or the direct participation of Mexican students in the collaborations working 
in the forthcoming upgrade to 12 GeV of JLAB. On our side we insisted in the need of non-conventional 
training programs for the first Mexican PhD students, who would require not only skills on 
simulations and design of components but also the experience of building, testing and operating them in 
order to have the knowledge to eventually lead groups on the corresponding topics.  

As a product of this collaboration a student got his Ph D (Alejandro Castilla Loeza) working on the 
design, building and testing of a crab cavity for the 12 GeV upgrade of JLAB; three  students got 
his/her Master degree (Karim Hern\'andez Chain, Celene Cuevas  and Christian L\'opez) and at least 
seven students participated in the JLAB summer program (Christian Valerio, Salvador Sosa, 
Luis Medina, Ernesto Barrientos, Sa\'ul Cuen, Cesar Serna y Gabriel Palacios up to 2013). Presently,
 Salvador Sosa is about to finish his Ph D program working at JLAB and another student 
 (Daniel Ch\'avez Valenzuela) is working on the design of the magnets for the upgrade JLAB 
 in a collaboration with Texas A\& M University and Guanajuato University.
 
Most of the work done by Mexican students at JLAB had finantial support from both, Mexican institutions 
and JLAB. However, unlike the CERN collaboration, CONACyT only contributed with the national grants and 
the "Becas Mixtas" program to this process and eventually the crisis in the USA reached 
scientific projects including those of JLAB making impossible to support further the work 
of Mexican students. Under this situation some students got their master at JLAB and continued working 
in the field at CERN under the auspices of the BEAM project.

\section{The Mexican Particle Accelerator School.}

As part of the initiatives to trigger the development of accelerator science and technology in Mexico 
by the ends of 2010 the then Deputy General Direction for Scientific Development of 
CONACYT, Dr. J. A. de la Pe\~na, created another panel with people interested in having a 
Mexican Synchrotron to work on several aspects considered important before starting formally 
the process, mainly a meeting of Mexican users of synchrotron facilities to have an idea of the 
size  of this community and the formulation of a methodological route for the process.
To this end the group received a grant from CONACyT leaded by one of us (M.N.). As part of the 
activities of this group we organised the First Mexican Particle Accelerator School (MEPAS I) on 
November 2011 at the city of Guanajuato. 

The program of MEPAS-I was designed by our colleagues at JLAB under the leadership of Carlos 
Hern\'andez and included 
talks by leaders of the accelerator divisions of several synchrotron facilities in the world and also 
some high energy physics facilities. In particular, it was very important the participation of 
Frank Zimmerman from CERN, which helped to close the loop of connections of the Mexican community 
with the accelerator international community. 

It took a long time to organise the second Mexican Particle accelerator School (MEPAS-2) which 
was held on November 2015 at Guanajuato city. However, during these four years between the 
first and the second edition, the number of students involved in the field grow up considerably 
as we will review in the next section, and in addition to leaders of several organisations 
(remarkably William Barletta director of the United States Particle Accelerator School, Liu Lin 
from the Brazilian Synchrotron  and Robert Hettel from SLAC) we had an active participation of the first 
Mexicans graduated in the field and PhD students working at JLAB, KEK, 
CERN and other facilities, as instructors. The second MEPAS had financial support mainly 
from the Red-FAE.

The third edition of MEPAS will be held this year under the auspices of the Mesoamerican Center 
for Theoretical Physics (MCTP) and will be completely organised by the Mexican Particle Accelerator Community of (CMAP).

\section{Mexican Particle Accelerator Community}

In addition to the JLAB and CERN initiatives during the past five years other institutions have 
been involved in the development of Mexican human resources in the field. One of them is the Texas A\& M 
Magnet Laboratory where a student (Daniel Ch\'avez Valenzuela ) got his Master and is presently 
pursuing hid Ph. D. working on the design of the magnets for the upgrade to 12 GeV of JLAB. Other 
students got involved in accelerator projects at ALBA, DESY, FERMILAB, ALS, TRIUMF and Liverpool 
University.

During the past two years Mexican students working at different laboratories in the world and 
the first Mexican PhD in the field have been discussing the need to focus efforts and as a first 
step they have organised in the Mexican Community of Particle Accelerators. This community has 
about 25 members and although most of them are graduate students working for their Ph D degree 
at several institutions, they are already committed to the conceptual design
and operation of the first Linear Accelerator (LINAC) built in Mexico \cite{Cristian2016,Humberto2016}.

\section{Accelerator Projects in Mexico}

There is no doubt that the project that triggered the whole process was the Mexican Synchrotron Light 
Source, which has been pushed by different groups in the past. After the original proposal for this 
light source for the CONACyT call for proposals of large scientific and technological projects in 
2006, the idea evolved with the grant to a different group in 2010 described above which boosted the 
training of students and lead to the formation of the Mexican Synchrotron Radiation Users Meeting which 
has had an active role in pushing the project. A new grant was provided by CONACyT and the state 
of Morelos in 2014 within the "Fondos Mixtos" program to a group which now included representatives 
of the user's community. The goal is to have  a conceptual design report for the Mexican Synchrotron 
Light Source this year aiming to construct it in the state of Morelos.
  
In parallel, some smaller projects involving accelerator science and technology have been 
proposed by different groups. Among them, an Electron Beam Irradiation Facility for 
"cold pasteurisation" of fruit has been pushed by the state of Michoac\'an and Michoac\'an State University (UMSNH); an Excellence 
Centre for Nuclear Medicine aiming to produce radio-pharmaceutical products for imagenology 
and cancer therapy has been proposed in Puebla and some applications on national security 
are being implemented in the state of Puebla. 
   
\section*{References}

\bibliography{AccMex}

\end{document}